\documentclass[review,3p,times,authoryear,11pt]{elsarticle}

\usepackage{tikz}
\usepackage{graphicx}
\usepackage{amssymb}
\usepackage{verbatim}
\usepackage{adjustbox}
\usepackage{algorithm}
\usepackage{algorithmic}
\usepackage{booktabs}
\usepackage{colortbl}
\usepackage{arydshln}
\usepackage{multirow}
\usepackage{rotating}
\usepackage{hyperref}
\usepackage{subfigure}
\usepackage{color}
\usepackage{clrscode3e}
\usepackage{longtable}
\usepackage{setspace}
\usepackage{threeparttable}
\usepackage[T1]{fontenc}	
\usepackage{chngcntr}
\counterwithout{figure}{section}
\usepackage{amsthm}
\usepackage{amsmath}
\usepackage{siunitx}

\usepackage{overpic}
\usepackage{graphicx,amsfonts,amssymb,amsmath,amsthm}

\allowdisplaybreaks[1] 
\newlength{\commentindent}
\usepackage{longtable}
\usepackage{graphicx}
\usepackage{algorithm}
\usepackage{algorithmic}
\usepackage{booktabs}
\usepackage{multirow}
\usepackage{graphicx}
\usepackage{float}
\usepackage{diagbox}
\usepackage{lscape}
\usepackage{xcolor}
\usepackage{geometry}
\usepackage{pgfplots}
\journal{}
 
\begin{document}

\begin{frontmatter}



\title{
The Multi-Trip Time-Dependent Mix Vehicle Routing Problem for Hybrid Autonomous Shared Delivery Location and Traditional Door-to-Door Delivery Modes
}


\author{Jingyi Zhao$^{1+}$}
\author{Jiayu Yang $^{3+}$}
\author{Haoxiang Yang$^{2*}$}

\address{$^1$ Shenzhen Research Institute of Big Data, Shenzhen 518172, China}
\address{$^2$ School of Data Science, The Chinese University of Hong Kong, Shenzhen (CUHK-Shenzhen), Shenzhen 518172, China}
\address{$^3$ Department of Mathematics, The University of Hong Kong}

\cortext[corl]{Corresponding Author;+Equal Contribution}

\begin{abstract}
Rising labor costs and increasing logistical demands are creating significant challenges for modern delivery systems. Automated Electric Vehicles (AEV) have emerged as a potential solution to reduce reliance on delivery personnel and increase route flexibility. However, the adoption of AEV is still limited due to varying levels of customer acceptance and the logistical complexities associated with integrating AEV into traditional delivery systems. Shared Distribution Locations (SDL) offer a viable alternative to door-to-door (D2D) distribution by providing a wider window of delivery time and the ability to serve multiple customers within a community. SDL has been shown to improve last-mile logistics by reducing delivery time, lowering costs, and increasing customer satisfaction.
In this context, this paper introduces the Multi-Trip Time-Dependent Hybrid Vehicle Routing Problem (MTTD-MVRP), 
a novel and particularly challenging variant of the VRP that incorporates Autonomous Electric Vehicles (AEVs) alongside conventional delivery vehicles. The difficulty arises from its unique combination of factors—time-dependent travel speeds, strict time windows, battery limitations, and driver labor constraints—while also integrating both SDL and IDL.
To tackle the MTTD-MVRP efficiently, we develop a tailored meta-heuristic based on Adaptive Large Neighborhood Search (ALNS), augmented with column generation (CG). This combined approach explores the solution space intensively through problem-specific operators and adaptively refines solutions, striking a balance between high-quality outcomes and computational effort. Extensive computational experiments show that the proposed method delivers near-optimal solutions even for large-scale instances within practical time limits.
From a managerial perspective, our findings underscore the importance of simultaneously considering autonomous and human-driven vehicles in last-mile logistics. Decision-makers can leverage the flexibility of routing through SDLs to reduce operational costs and carbon footprints, while still accommodating customers who require or prefer door-to-door services.


\end{abstract}


\end{frontmatter}

\section{Introduction} \label{Section:Introduction}
In Asian countries, especially China, the conflict between rising labor costs and the growing demand for social logistics is becoming increasingly apparent \citep{supplychainasia2024autonomous}. Dr. Kong Qi, chief scientist and head of JINGDONG (JD) autonomous driving technology, suggests that China's advancements in Autonomous Delivery Vehicles (ADV) will help tackle this issue since ADV can reduce the reliance on delivery staff and allow for more flexible delivery routes.  
Not coincidentally, companies like Tesla have introduced autonomous electric vehicles (AEV) to the passenger car market, while technology firms like Google and Amazon have expanded into last-mile logistics with drones. 
In recent years, industry players like China's Cainiao and Meituan have also entered this market \citep{BinghuW2023}. 
For instance, in practice, Meituan's intelligent distribution system assigns grocery orders to AEV, which picks up the orders from designated sites and autonomously delivers them to contactless distribution points in residential communities, where recipients collect their items without human contact \citep{Logclub2024}. 
The "human-machine collaboration" approach proposed by Meituan, involving both AEV and courier services, is currently being used in nearly 30 cities in China to provide last-mile delivery. 
Their statistics show that the increase in AEV during holidays and promotional events resulted in a 353\% year-on-year increase in completed delivery orders \citep{CN1562024}.

However, many people have not yet adapted to this mode of automated delivery, and some elderly and disabled still require manual assistance.
While younger customers are more receptive to this new AEV delivery method, they can usually only take over the service on weekends or weekday evenings.
This is because customers must be home and pick up their goods when AEV deliveries are made to their door.
So customers will want to receive their packages within a specific time window, which limits the company's flexibility in designing routes, and weekend and weekday evenings can be overloaded with tasks for which AEV may not be sufficient.
Therefore, a better option is combining AEV with shared delivery locations (SDL), such as supermarkets or security offices. 
The rise of SDL as an alternative to traditional door-to-door (D2D) delivery marks a significant shift in urban logistics \citep{AmazonKey2024}. 

Besides the companies, the researchers in the VRP field have also observed that SDL enhances the efficiency of last-mile distribution systems, improving the quality of designed routes \citep{Drexl2020,Tilk2021}. 
It offers broader delivery time windows and the ability to serve multiple customers within a community, reducing overall delivery time, lowering costs, and providing greater customer flexibility. 
As a result, SDL has gained popularity among office workers across Asia, and the successful application of autonomous vehicles in some regions suggests a trend toward broader adoption. However, acceptance of autonomous vehicles and SDL varies, as some customers still prefer traditional door-to-door service due to concerns about data privacy. Therefore, we propose a more flexible delivery model that allows customers to choose between AEV-SDL and traditional manned door-to-door delivery.

\begin{figure}[htbp]
\centering
\includegraphics[width=1\textwidth]{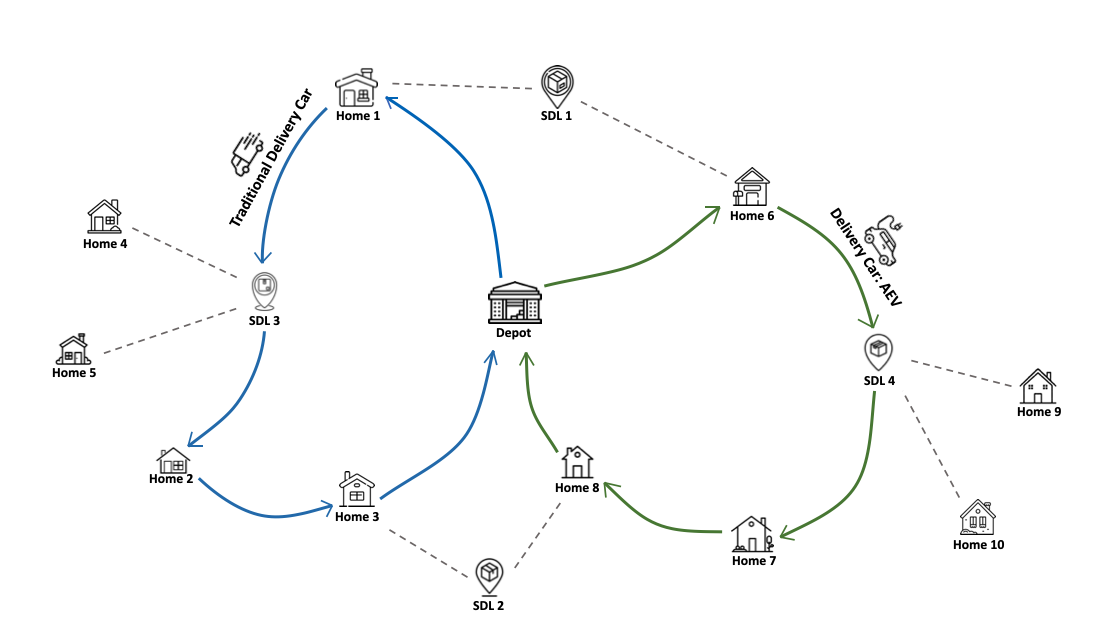}
\caption{An example of a delivery plan for MTTD-MVRP.}
\label{exam}
\centering
\end{figure}
In this context, we introduce a complex multi-trip time-dependent hybrid vehicle routing problem (MTTD-MVRP) that is as realistic as possible, which addresses the logistical challenges posed by the different operational parameters of self-driving and human-driven vehicles, as well as by the co-existence of SDL and traditional Individual Distribution Locations (IDL). The model can cope with complex time-varying travel speeds, customer service time windows, and vehicle-specific constraints, including drive time for conventional vehicles and battery limitations for AEVs.
Fig.~\ref{exam} illustrates the interaction between vehicles and various types of distribution locations, including SDLs and IDLs. In this example, customers 4 and 5 share SDL 3, which is located on the route between customers 3 and 2. This setup provides a better alternative to a traditional delivery vehicle, allowing it to fulfill orders for customers 4 and 5 without actually reaching those homes. Although AEVs can only serve customers who choose this option, realistically, AEVs can be used whenever a customer prefers to use the SDL pickup option.
We aim to design a routing strategy that minimizes the total operating cost while ensuring customer satisfaction and operational feasibility.
To this end, we propose a tailored meta-heuristic Adaptive Large Neighbourhood Search (ALNS) with Local Search (LS) 
augmented with column generation (CG) and equipped with newly designed problem-specific operators to solve this problem efficiently.
Our contribution is threefold:

\begin{enumerate}
    \item We introduce a new problem, MTTD-MVRP, which addresses the integration of conventional and autonomous electric vehicles. By incorporating elements of customer preferences for self-driving car services and shared pick-up points, the proposed model provides a comprehensive framework for urban logistics that is both efficient and considers the development of emerging technologies. 
We propose a self-consist mathematical model for this problem that captures each vehicle's unique operating characteristics and constraints, providing a more comprehensive and realistic model for urban logistics.

\item We propose a meta-heuristic framework ALNS-LS with problem-specific designed operators for the MT-TDMVRP. Additionally, introducing a Markov chain for the dynamic selection of Local Search (LS) operators accounts for each LS operator's enhancing and diminishing effects. This approach accelerates the convergence of the refined solution and increases the likelihood of uncovering a superior solution. Traditional roulette wheel selection methods often fail to consider the interaction effects between operators, leading to suboptimal outcomes. Our heuristic features a unique set of operators and a tailored adaptation process strategically integrated to facilitate efficient solution space exploration.
\item Our method has demonstrated remarkable results, identifying an unprecedented number of optimal solutions across a spectrum of data instances. This achievement is particularly notable in the Duration-Minimizing Time-Dependent Vehicle Routing Problem with Time Windows (DM-TDVRPTW), where our approach has significantly improved solution quality and computational efficiency. When compared to the hybrid method proposed by \cite{pan2021hybrid}, our ALNS-LS algorithm exhibits superior performance, solidifying its position as a leading algorithm for solving complex routing and inventory problems.

\end{enumerate}

\section{Related studies} \label{sectionre}
\noindent
\textbf{The time-dependent vehicle routing problem (TDVRP)}
was first introduced in \cite{malandrakiDaskin1992td_f}, where a stepwise travel time model was employed. Subsequently, \cite{Ichoua2003} proposed an alternative approach, representing travel speeds with a stepwise function, leading to a piecewise linear travel time function. Building upon this model, many studies have investigated different heuristic algorithms to address variants of TDVRP with time windows (TDVRPTW). Some notable approaches include the applications of ant colony systems \citep{Donati2008td_ant, Balseiro2011, liu2020time}, genetic algorithm \citep{haghani2005dynamic,zhao2024hybrid}, and adaptive large neighborhood search algorithm \citep{zhang2020time, pan2021multi,zhao2022adaptive}. Different from TDVRPTW, the duration-minimizing TDVRPTW (DM-TDVRPTW) focuses on minimizing the total duration of routes, which introduces the extra decision on the departure time at the depot and thereby dramatically increases the computational complexity \citep{Dabia2013bp_tdvrptw, pan2021hybrid}.
For a comprehensive review of TDVRP and its variants, interested readers may refer to \cite{Gendreau2015TD_review}.\\

\noindent
\textbf{The vehicle routing problems with alternative delivery locations (VRPDO)} have been studied in recent years, aiming to address last-mile logistics challenges. In this type of problem, each order is associated with a set of alternative delivery locations, including individual addresses and shared delivery points. Some researchers, such as \cite{reyes2017vehicle}, considered scenarios where customers provide alternative delivery locations, each with a specified time window for delivery (e.g., home, office, or gym). Others, like \cite{ozbaygin2017branch}, explored variants involving both home and roaming delivery locations and designed the branch-and-price algorithm to get the optimal solutions. \cite{sitek2019capacitated} addressed a similar problem but ignored the time window constraints. Recent contributions on SDL include works by \cite{orenstein2019flexible,janjevic2019integrating,enthoven2020two}. However, these studies focused solely on the diversity of customer delivery locations, overlooking the degree of priority among several optional delivery locations. This situation is addressed by \cite{dumez2021large}, who considered the priority level of different options and required the service level of chosen options to satisfy a specific minimum level. They devised a matheuristic approach but only solved the set-partitioning model built based on the routes obtained by the large neighborhood search to get a new solution. \cite{tilk2021last} addressed a similar problem to \cite{dumez2021large} and proposed a branch-and-price-and-cut algorithm. In addition, some studies did not explicitly model the priorities regarding different delivery options but implicitly reflect the preference with the specific compensation or cost \citep{baldacci2017vehicle,zhou2018multi,grabenschweiger2021vehicle}. Lastly, the Generalized Vehicle Routing Problem with Time Windows is a special case of the VRPDO but with a single priority level. In this case, customer requests can be satisfied at various locations, but SDL is not considered \citep{moccia2012incremental,vidal2014unified,yuan2021column}.\\

\noindent 
\textbf{The autonomous electric vehicle (AEV) delivery services} have emerged as a prominent area of research within the logistics industry. Integrating AEV into delivery services can revolutionize urban logistics by offering a more sustainable, efficient, and cost-effective alternative to traditional vehicle fleets. 
Research has demonstrated that AEV can significantly reduce labor costs associated with delivery operations and enhance safety by minimizing the risk of human error during transit \citep{zhang2020}. 
AEV is particularly effective in scenarios with repetitive tasks and predictable environments, such as last-mile delivery in urban areas. They can operate around the clock with minimal breaks, thereby increasing the overall efficiency of delivery operations.
However, the deployment of AEV is not without challenges. Technical limitations, such as the need for precise navigation systems and the ability to handle complex traffic situations, remain significant barriers to widespread adoption \citep{li2021}.
Additionally, regulatory frameworks are still being developed, with legal standards for AEV operation varying greatly across different jurisdictions. This creates a complex landscape for logistics companies looking to integrate AEV into their service offerings.
The development of AEV technology is also closely tied to advancements in areas such as battery life, charging infrastructure, and vehicle-to-everything communication. As these technologies mature, AEV is expected to become increasingly capable of handling a broader range of delivery tasks \citep{santini2018comparison}.

In conclusion, while self-driving AEV offers many advantages for delivery services, their full potential remains to be realized. Our research and development efforts hope to better integrate current realities, i.e., with the types of services that people are currently accustomed to and accept, to facilitate the seamless integration of AEV into urban logistics.
Moreover, although some studies on VRPDO have explored the diversity of delivery locations, they overlooked the combined complexities of time-dependent (TD) factors and the integration of hybrid delivery vehicles. 
On the other hand, studies on the AEV routing problem have demonstrated the potential of hybrid distribution vehicles, but these studies fail to simulate real-world scenarios by ignoring considerations such as time windows and TD road conditions.
To bridge these gaps, we creatively integrate SDL and AEV while incorporating time windows and different speed profiles. This intends to provide a robust model that aligns more with the dynamic nature of modern urban logistics.
\section{Problem description and modelling} \label{section:prob_desc}

 Our MTTD-MVRP is formulated on a complete graph \( G = (\mathcal{V}, \mathcal{A}) \), where \( \mathcal{V} \) represents the set of nodes \(\{0, 1, \ldots, 2n+1\}\), and \( \mathcal{A} \) denotes the set of arcs. The set of nodes \( \mathcal{V} \) includes the origin depot 0, the destination depot \( 2n+1 \), the delivery nodes (representing the residences of customers) \( \mathcal{V}_{d} = \{1, 2, \ldots, n\} \), and the SDL nodes \( \mathcal{V}_{s} = \{n+1, n+2, \ldots, 2n\} \).
Each SDL node $(i + n)$ is paired with its corresponding delivery node $i$, two SDL nodes may share the same location, reflecting a scenario where multiple customers from the same neighborhood or village typically use a common SDL. The set of arcs \( \mathcal{\mathcal{A}} \) is defined as \( \{(i, j) \mid i \in \mathcal{\mathcal{V}} \setminus \{2n+1\}, j \in \mathcal{\mathcal{V}} \setminus \{0\}, i \neq j \} \), and each arc \( (i, j) \) is associated with a distance \( d_{i, j} \).
The MTTD-MVRP problem aims to minimize the total time and operational costs of all vehicles, considering multi-type vehicles, time-dependent speed, time windows, and customer preferences.
We consider two types of vehicles: 1)  delivery fuel vehicles and 2) AEVs. 
Specifically, two groups consisting of \( C_{1} \) number of traditional delivery vehicles and \( C_{2} \) number of AEVs with load capacities \( Q_{1} \) and \( Q_{2} \) are based at the depot. Fuel vehicle routes are constrained by the maximum working time \( H_{1} \) of drivers, while AEV routes are limited by the maximum driving distance \( P_{2} \), reflecting restrictions imposed by their battery capacity.

In this work, we consider a total of \( n \) customers awaiting delivery, each with the option to accept or decline AEV and SDL services. 
These preferences are represented by binary variables \( A_{i} \in \{0, 1\} \) and \( S_{i} \in \{0, 1\} \) for each customer \( i \in \{1, \ldots, n\} \). 
Specifically, \( A_{i} = 1 \) signifies that the customer \( i \) consents to use AEV for their delivery, and \( S_{i} = 1 \) indicates that they consent to their packages being stored at SDL.
Each customer \( i \) has a specified delivery demand \( D_{i} \) scheduled for delivery to the corresponding delivery node (\( i \in \mathcal{V}_{d} \)) within a strict time window $[e_{i}, l_{i}]$. 
Customers who accept the SDL service, indicated by $S_{i} = 1$, can collect their delivery $L_{i}$ from the SDL node ($i + n \in \mathcal{V}_{s}$) within a generally wider and flexible time window $[e_{i+n}, l_{i+n}]$. 

The operational cost for a vehicle's route is calculated as the total time elapsed from the vehicle's departure from the depot to its return to the destination depot, plus the fixed costs incurred for each activated vehicle, which depends on the vehicle type and the total number of vehicles activated.
\subsection{Time-dependent travel speed function}
The MTTD-MVRP model utilizes the time-dependent travel time approach proposed in \cite{Ichoua2003}, which holds the first-in-first-out (FIFO) principle. This model divides the maximum time window \([e_{0},l_{2n+1}]\) into different time steps that do not intersect. \(p\) Each of them is assigned a constant value representing the vehicle's speed during that specific period. This setup, applied to each edge \((i,j) \in \mathcal{\mathcal{A}}\), is referred to as the speed profile of the edge. Given this piecewise constant time-speed relationship, the resulting function that maps the departure time from \(i\) to the travel time across the edge becomes a piecewise linear function. Consequently, all information about this function can be represented and stored using a maximum of \(2p\) breakpoints (including 0, with some breakpoints possibly overlapping or falling outside the timeline, resulting in fewer breakpoints), generated by time zones \(p\) (\cite{Ichoua2003}; \cite{pan2021multi}). We then use an indexed set \(T_{i,j} = \{0,1,...,|T_{i,j}|-1\}\) to enumerate each time step split by breakpoints \(w^{0}_{i,j}, w^{1}_{i,j},...,w^{|T_{i,j}|-1}_{i,j}\) for each edge \((i,j)\), and then define the \(k^{th}\) time step as \(\Delta T_{i,j}^{k}=[w_{i,j}^{k},w_{i,j}^{k+1})\).

\begin{figure}[htbp]
\centering
\includegraphics[width =17cm]{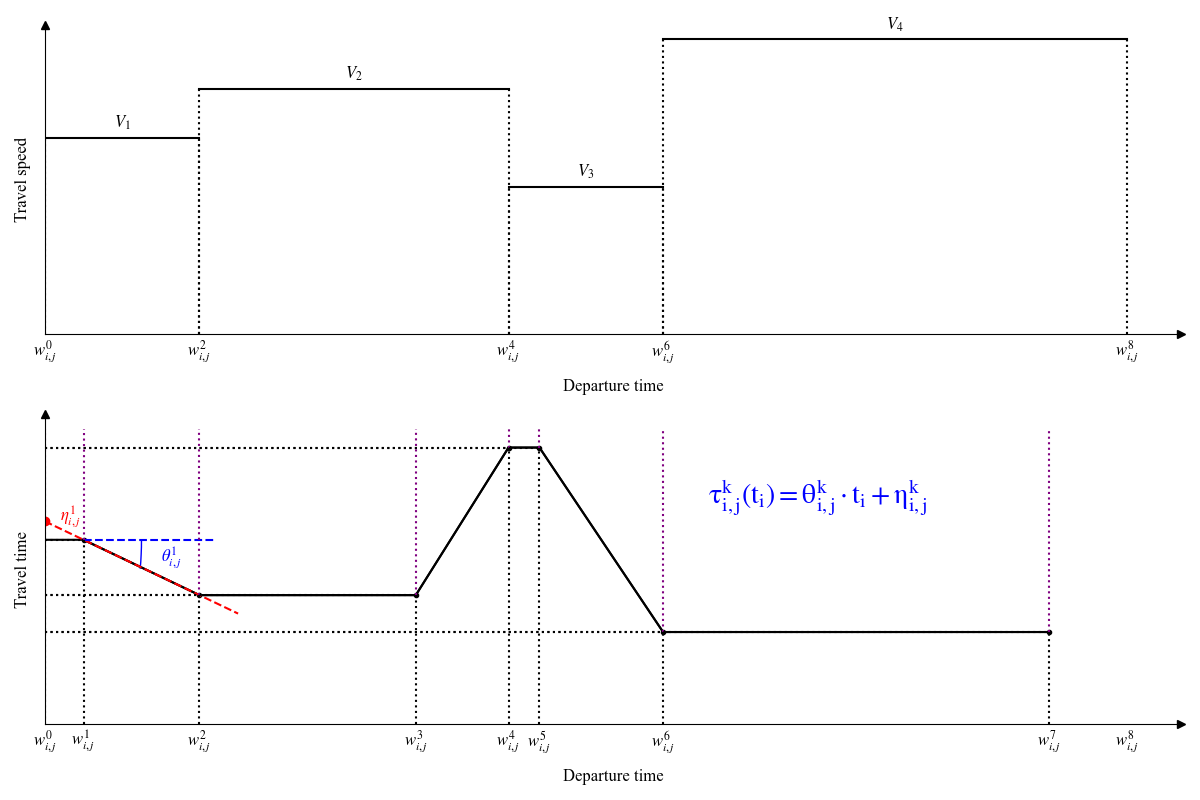}
\caption{Example of Time-dependent Travel speed and travel time function}
\label{Time-dependent travel speed function}
\end{figure}

Let \(\tau_{i,j}\) denote the piece-wise linear function mapping from departure time to travel time for the edge \((i,j)\). We define the linear function \(\tau_{i,j}^{k}\) by constraining \(\tau_{i,j}\) within \(\Delta T_{i,j}^{k}\) (\(k \in T_{i,j}\)). The slope \(\theta_{i,j}^{k}\) and y-intercept \(\eta_{i,j}^{k}\) of this function can be calculated from nearby successive breakpoints $\bigg(w_{k},\tau_{i,j}(w_{k})\bigg)$ and $\bigg(w_{i,j}^{k+1},\tau_{i,j} (w_{i,j}^{k+1})\bigg)$.

The time-dependent travel speed function is defined as follows:
\begin{equation}
    \tau_{i,j}^{k}(t_{i}) = \theta_{i,j}^{k} \cdot t_{i} + \eta_{i,j}^{k}, \quad \forall t_{i} \in \Delta T_{i,j}^{k}
\end{equation}

We enumerate type 1 fuel vehicles as \( 1, 2, \dots, M_1 \in C_{1} \) and type 2 AEVs as \( 1, 2, \dots, M_2 \in C_{2} \), respectively.  Binary variables \( x_{i,j}^{v,m,k} \) are defined to indicate whether the \( m^{\text{th}} \) vehicle of type \( v \) (\( v = 1 \): fuel vehicle; \( v = 2 \): AEV) traverses the arc \((i,j)\) with a departure time within \(\Delta T_{i,j}^{k}\). The variable \((m\) represents the index of this type of vehicle within \(m \in \mathcal{C}_{v}\). Furthermore,  \(t_{i,j}^{v,m,k}\) indicate the departure time of the \(m^{th}\in C_{v}\) vehicle of type \(v \in \{1,2\}\) vehicle from node \(i\) to node \(j\). If the vehicle never visits the edge \((i,j)\) within \(\Delta T_{i,j}^{k}\), the variable \(t_{i,j}^{v,m,k}\) is set to 0.

\subsection{Mathematical formulation in mixed integer linear programming}\label{MILP}
 We define the decision variables as follows:

- \(x_{i,j}^{v,m,k}\) (\((i,j) \in \mathcal{A}, v \in \{1,2\}, m \in C_{v}, k \in T_{i,j}\)) is a binary variable set to one if the \(m^{th}\) vehicle of type \(v\) departs from node \(i\) to node \(j\) within the time period \(\Delta T_{i,j}^{k}\), and 0 otherwise.
  
- \(t_{i,j}^{v,m,k}\) is a non-negative continuous variable representing the departure time of the \(m^{th}\) vehicle of type \(v\) when traveling from node \(i\) to node \(j\) within the time period \(\Delta T_{i,j}^{k}\), and 0 otherwise.
  
- \(\gamma_{i}^{v,m}\) is a non-negative continuous variable indicating the departure time of the \(m^{th}\) vehicle of type \(v\) from node \(i\). If the vehicle never visits node \(i\), the variable is set to 0.
  
- \(q_{i}^{v,m}\) is a non-negative continuous variable denoting the load of the \(m^{th}\) vehicle of type \(v\) after serving node \(i\). If the vehicle never visits node \(i\), the variable is set to 0.
  
- \(d_{i}^{m}\) is a non-negative continuous variable representing the travel distance of the \(m^{th}\) type 2 vehicle: AEV after leaving node \(i\). If the vehicle never visits node \(i\), the variable is set to 0.

- \(F_{v}\) represents the fixed operational cost for using a type \(v\) vehicle.

- \(M\) denotes a sufficiently large number, which serves as a proxy for \(+\infty\) in the equations.

We present the MILP model of MTTD-MVRP as follows:

\begin{equation}
    \text{min} \sum_{v \in \{1,2\}} \sum_{m=0}^{\mathcal{C}_{v}} \left((\gamma_{2n+1}^{v,m} - \gamma_{0}^{v,m}) + (1 - \sum_{k \in T_{0,2m+1}}x_{0,2m+1}^{v,m,k})F_{v}\right)\label{obj}
\end{equation}
\begin{align}
\text{s.t.} 
    &\sum_{j \in \mathcal{V}\setminus \{0\}} \sum_{k \in T_{0,j}} x_{0,j}^{v,m,k} = 1, \qquad \forall v \in \{1,2\}, m \in \mathcal{C}_{v} \label{origin_out} \\
    &\sum_{j \in \mathcal{V}\setminus \{2n+1\}} \sum_{k \in T_{i,2n+1}} x_{i,2n+1}^{v,m,k} = 1, \qquad \forall v \in \{1,2\}, m \in \mathcal{C}_{v} \label{destination_in} \\
    &\gamma_{i}^{v,m} = \sum_{j\in \mathcal{V}\setminus\{0\}} \sum_{k \in T_{i,j}} t_{i,j}^{v,m,k}, \qquad \forall v \in \{1,2\}, m \in \mathcal{C}_{v} \label{time_consistency} \\
    &\sum_{j \in \mathcal{V}\setminus \{0,2n+1\}} \sum_{k \in T_{i,j}} x_{i,j}^{v,m,k} =\sum_{h \in \mathcal{V}\setminus \{0,2n+1\}} \sum_{k \in T_{h,i}} x_{h,i}^{v,m,k}, \qquad \forall v \in \{1,2\}, m \in \mathcal{C}_{v}, i \in \mathcal{V}\setminus\{0,2n+1\} \label{flow_in_out} \\
    &\sum_{i \in \mathcal{V} \setminus \{0\}} \sum_{k \in T_{i,j}} x_{i,j}^{v,m,k} + \sum_{i \in \mathcal{V} \setminus \{0\}} \sum_{k \in T_{i,j+m}} x_{i,j+m}^{v,m,k} = 1, \qquad \forall v \in \{1,2\}, m \in \mathcal{C}_{v}, j \in \mathcal{V}_{d} \label{Nc_Ns} \\
    &\gamma_{j}^{v,m} \ge \gamma_{i}^{v,m}(1+\theta_{i,j}^{k})x_{i,j}^{v,n,k} + (\eta_{i,j}^{k}+s_{j})x_{i,j}^{v,n,k}, \qquad \forall i \in \mathcal{V}\setminus\{2n+1\}, j \in \mathcal{V}\setminus\{0\}, v \in \{1,2\}, m \in \mathcal{C}_{v}, k \in T_{i,j} \label{time_progress} \\
    &w_{k}x_{i,j}^{v,m,k} \le t_{i,j}^{v,m,k} \le w_{k+1}x_{i,j}^{v,m,k}, \qquad \forall i \in \mathcal{V}\setminus\{2n+1\}, j \in \mathcal{V}\setminus\{0\}, v \in \{1,2\}, m \in \mathcal{C}_{v},\ k, k+1 \in T_{i,j} \label{time_interval} \\
    &q_{i}^{v,m}+{D}_{j} \le q_{j}^{v,m} + M(1-\sum_{k \in T_{i,j}} x_{i,j}^{v,m,k}), \qquad \forall i \in \mathcal{V}\setminus\{2n+1\}, j \in \mathcal{V}\setminus\{0\}, v \in \{1,2\}, m \in \mathcal{C}_{v} \label{load_progress} \\
    &\gamma_{j}^{v,m} \le (l_{j} + s_{j})\sum_{i \in N\setminus{2n+1}} \sum_{k \in T_{i,j}} x_{i,j}^{v,m,k}, \qquad \forall j \in \mathcal{V}\setminus\{0,2n+1\}, v \in \{1,2\}, m \in \mathcal{C}_{v} \label{time_window_1} \\
    &\gamma_{j}^{v,m} \ge (e_{j} + s_{j}) \sum_{i \in N\setminus{2n+1}} \sum_{k \in T_{i,j}} x_{i,j}^{v,m,k}, \qquad \forall j \in \mathcal{V}\setminus\{0,2n+1\}, v \in \{1,2\}, m \in \mathcal{C}_{v} \label{time_window_2} \\
    &q_{2n+1}^{v,m} - q_{0}^{v,m} \le Q_{v}, \qquad \forall v \in \{1,2\}, m \in \mathcal{C}_{v} \label{max_load} \\
    &\gamma_{2n+1}^{1,m} - \gamma_{0}^{1,m} \le H_{1}, \qquad \forall m \in \mathcal{C}_{1} \label{working_hour} \\
    &d_{j}^{m} \ge d_{i}^{m} + dis(i,j) - \mathrm{M}(1-\sum_{k \in T_{i,j}} x_{i,j}^{2,m,k}), \qquad \forall i \in \mathcal{V}\setminus\{2n+1\}, j \in \mathcal{V}\setminus\{0\}, m \in \mathcal{C}_{2} \label{dis_progress}\\
    &d_{2n+1}^{m} - d_{0}^{m} \le P_{2}, \qquad \forall m \in \mathcal{C}_{2} \label{max_dis} \\
    &\sum_{i \in \mathcal{V} \setminus {2n+1}} \sum_{k \in T_{i,j}} x_{i,j}^{v,m,k} \le S_{j}, \qquad \forall j \in \mathcal{V}_{s}, v \in \{1,2\}, m \in \mathcal{C}_{v} \label{self-pickup} \\
    &\sum_{i \in \mathcal{V} \setminus {2n+1}} \sum_{k \in T_{i,j}} x_{i,j}^{2,m,k} +\sum_{i \in \mathcal{V} \setminus {2n+1}} \sum_{k \in T_{i,j+n}} x_{i,j+n}^{2,m,k} \le A_{j}, \qquad \forall m \in \mathcal{C}_{2}, j \in \mathcal{V}_{d} \label{unmanded-vehicle}
\end{align}

We aim to minimize the objective function, which represents the total time taken by all vehicles, along with the fixed costs associated with each used vehicle, as defined in Equation \eqref{obj}. Equations \eqref{origin_out} and \eqref{destination_in} ensure that the flow out of the origin and the flow into the destination are equal to one. Constraint \eqref{flow_in_out} maintains the consistency between the flow into and out of any node in \(\mathcal{V}_{d} \cup \mathcal{V}_{s}\). Constraint \eqref{Nc_Ns} ensures that any delivery node $i$ (\(i \in \mathcal{V}_{d}\)) and the corresponding SDL node $i+n$ (\(i+n \in \mathcal{V}_{s}\)) are visited exactly once in total. Meanwhile, \eqref{time_progress} guarantees that the departure time of any node must be greater than the departure time of the previous node plus the driving time and service time. The constraint \eqref{time_interval} constrains the departure time of any node within its corresponding time interval. Equation \eqref{load_progress} ensures that the current load is greater than the load in the previous node plus the amount of cargo in the current node. Constraints \eqref{time_window_1} and \eqref{time_window_2} guarantee the departure time is within the time window. Constraints \eqref{max_load}, \eqref{working_hour}, and \eqref{max_dis} ensure that the load and working hours of the vehicles do not exceed the maximum settings. Equation \eqref{dis_progress} is similar to \eqref{load_progress} but concerning distance. Constraint \eqref{self-pickup} checks whether customer \(i\) accepts the SDL service; if not, the corresponding SDL node \(i+n\) cannot be visited. Meanwhile, constraints \eqref{unmanded-vehicle} guarantee the customer accepts AEVs; if not, both the delivery node and the SDL node cannot be visited by any AEVs.
\section{Meta-heuristic Algorithm}
In this section, we introduce a hybrid metaheuristic algorithm developed to tackle the MTTD-MVRP. Our approach combines the ALNS and LS framework, incorporating problem-specific operators tailored to both LS and ALNS mechanisms. This integration significantly enhances the algorithm's effectiveness in addressing the complexities of the MTTD-MVRP. Recent studies have demonstrated the superiority of combining these two methods, as they complement each other: LS thoroughly explores local regions to find optimal solutions, while ALNS escapes local optima by exploring broader solution spaces. Furthermore, our algorithm integrates classical heuristic methods such as Simulated Annealing (SA) and Ant Colony Optimization (ACO), along with various self-adaptive mechanisms into the ALNS-LS framework. This design is to balance the trade-off between effectiveness and efficiency, ensuring that the algorithm can achieve high-quality solutions in a computationally efficient manner.
\par
\begin{figure}[htbp]
\centering
\includegraphics[width =10cm]{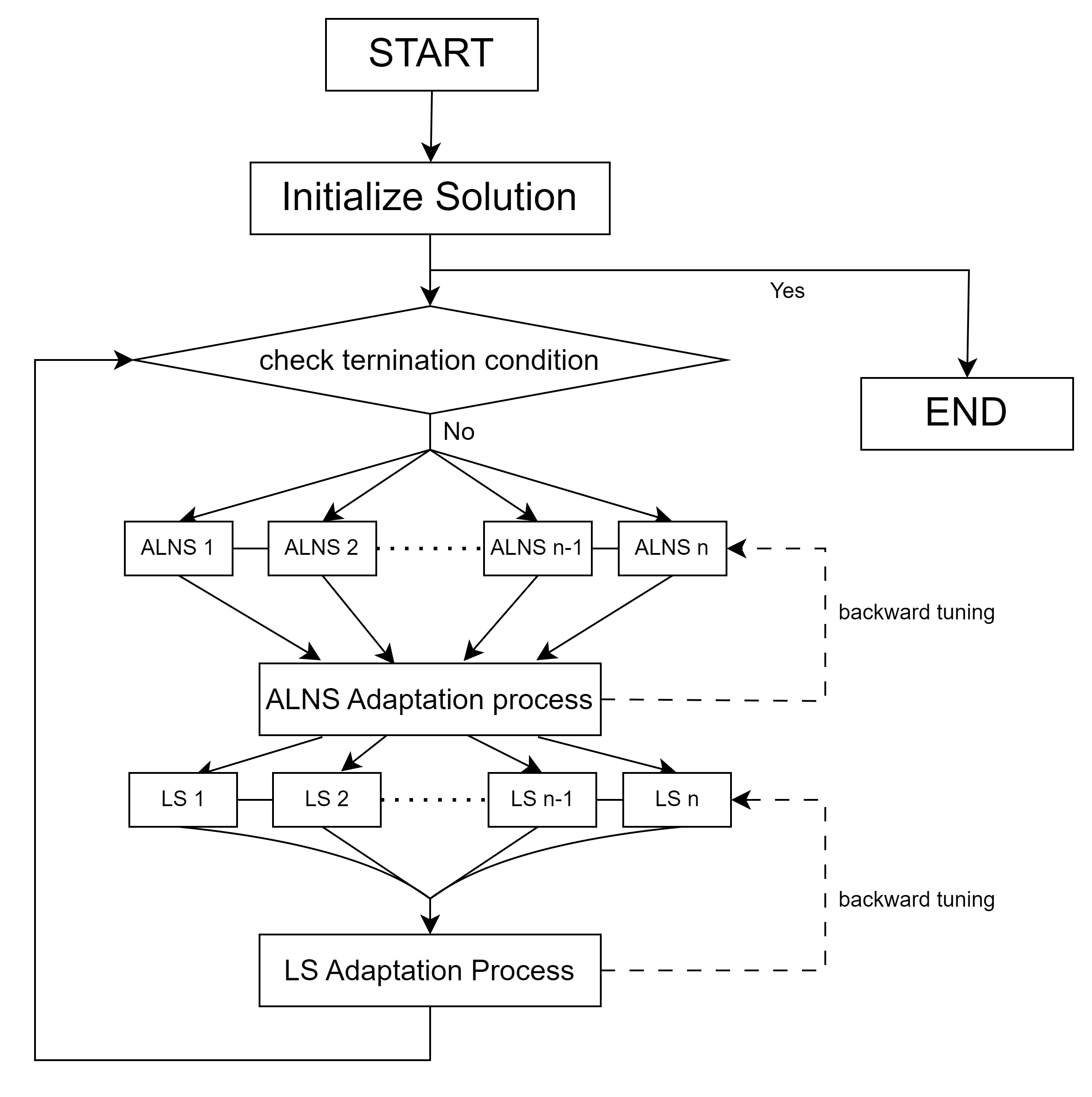}
\caption{Framework of hybrid meta-heuristic algorithm}
\label{meta_heuristic_flow_chart}
\end{figure}

 The general outline of our hybrid meta-heuristic algorithm for solving the MTTD-MVRP is shown in Figure \ref{meta_heuristic_flow_chart}. Initially, the algorithm constructs an initial solution using greedy reinsertion (section \ref{greedy reinsertion}). Then entering the ALNS iteration phase, the strength of the ruin and reconstruction process is determined in the adaptive process (section \ref{Dynamic perturbation strength}). We employ multi-threading techniques to generate multiple potential solutions concurrently for each ALNS iteration. This parallel approach not only makes the algorithm more hardware-efficient but also encourages the ALNS to explore the solution space more thoroughly, avoiding potential misses and increasing the likelihood of identifying high-quality solutions for subsequent processing. After completing ALNS, we proceed to an adaptive process. This adaptation is based on the performance of each generated solution in terms of objective cost and diversity, allowing the algorithm to dynamically adjust its strategies to prefer or reject any specific operators in ALNS. The solutions are rated, and those meeting certain criteria are passed on to the LS phase for further refinement. During the LS process, filtered solutions are further refined, followed by a similar adaptive process applied. This iterative process of ALNS and LS continues until the final termination condition is met.

\subsection{Initial solution}
\par To generate the initial solution, we mainly use the greedy method. Firstly, consider all nodes that have been removed and are awaiting reinsertion. Using greedy reinsertion operators (section \ref{greedy reinsertion}), we iteratively insert one node from the waiting list at a time until a feasible solution is obtained or until reinsertion is no longer possible within the partially constructed tour. If the latter occurs, a randomly chosen destroy operator (section \ref{Removal operators}) removes half of the nodes, after which the greedy reinsertion process is repeated. This process continues until a feasible solution is found.

\subsection{Route evaluation procure}
\par The feasibility check algorithm for a vehicle route is a crucial part of the heuristic approach. This algorithm is invoked numerous times during each iteration, in both the LS and ALNS processes. The overall efficiency of the algorithm can be significantly improved by efficiently designing the feasibility check procedure.
In this section, we present the design of the feasibility-checking evaluation method for the MTTD-MVRP. Ensuring a feasible route for the MTTD-MVRP requires satisfying the following requirements:

\begin{enumerate}
    \item The load volume of every type of vehicle must not exceed their respective capacity \(Q_{1}, Q_{2}\).
    \item Customer preferences must be strictly followed: If $S_{i} = 0$, the vehicle is prohibited from visiting node $i + n \in \mathcal{V}_{s}$. Similarly, if $U_{i} = 0$, any AEV is not permitted to visit nodes $i \in \mathcal{V}_{d}$ and $i + n \in \mathcal{V}_{s}$.
    \item Each visited node must be arrived at within its specified time window (arrival $\rightarrow$ serving $\rightarrow$ departure). Additionally, the working hours of type 1 traditional vehicles should not exceed the maximum working hours \(H_{1}\).
    \item  Each delivery node \( i \in \mathcal{V}_{d} \) and its corresponding SDL node \( i + n \in \mathcal{V}_{s} \) can be visited at most once in total within a single route.
    \item For AEV, the travel distance must not exceed the maximum working distance \(D_{2}\).
\end{enumerate}

\par Requirements 1, 2, and 4 are node-based and can be verified by straightforwardly examining the nodes included within the route. Requirement 5 can be evaluated through a sequential traverse of all nodes along the route. Verification of Requirement 3, however,  requires knowledge of the departure time from depot 0. As this time is flexible and influences arrival times at each node as well as the overall duration of the route, its determination typically involves the application of forward-propagation and back-propagation methods, as proposed by \cite{zhao2022adaptive}, with necessary modifications tailored to this problem. For a detailed explanation, refer to algorithm \ref{temporal feasibility evaluation} for the pseudo-code outlining this process.

\begin{table}[htp]
\centering
\caption{Notations in temporal feasibility evaluation}
\resizebox{0.95\textwidth}{!}{
\begin{tabular}{ll} 
\toprule
Symbol & Definition \\ 
\hline
$[\mathrm{ETW}_{S_{N}}, \mathrm{LTW}_{S_{N}}]$ & The earliest time window and latest time window of departure time from the node $S_{N}$ \\
$\mathrm{BPS}_{S_{N}\rightarrow S_{N+1}}$ & The original time breakpoints of node $S_{N}$ generated by time-dependent route $(S_{N}\rightarrow S_{N+1})$ \\
$\mathrm{W}_{S_{N}}$ & The serving time for unloading at node $S_{N}$ \\
$f_{S_{N} \rightarrow S_{N+1}}$ & The function returns the earliest possible departure time from node \( S_{N+1} \), given the departure time from node \( S_{N} \). \\
& $f_{S_{N} \rightarrow S_{N+1}}(t) = \tau_{ij}(t) + \mathrm{W}_{S_{N+1}} + t \quad t \in T_{S_{N},S_{N+1}}^{k}$ \\
& $f_{S_{N} \rightarrow S_{N+1}}^{-1}(t) = \frac{t - \mathrm{W}_{S_{N+1}} - \eta^{k}_{(S_{N},S_{N+1})}}{1 + \theta^{k}_{(S_{N},S_{N+1})}} \quad \,\,\,\, t \in f_{S_{N} \rightarrow S_{N+1}}(T_{(S_{N},S_{N+1})}^{k})$ \\
& $f_{S_{N} \rightarrow S_{N+1}}(T) = \{f_{S_{N} \rightarrow S_{N+1}}(t)\ |\ \forall t \in T\}$ \\
\bottomrule
\end{tabular}
}
\end{table}

\begin{algorithm}[htp]
\caption{Temporal Feasibility Evaluation}
\label{temporal feasibility evaluation}
\begin{algorithmic}[1]
\STATE Input: Route $S = (S_{0},S_{1}...S_{v-1},S_{v})$ with $S_{v} = 2n+1$ and $S_{0} = 0$
\STATE LTW $\leftarrow +\mathrm{M}$
\STATE ETW $\leftarrow 0$
\STATE \textbf{backward propagation}
\FOR{$\mathrm{node}\ S_{N} \in (S_{v-1},...,S_{1},S_{0})$}
\STATE LTW $\leftarrow \min(f_{S_{N} \rightarrow S_{N+1}}^{-1}(\mathrm{LTW}),\mathrm{LTW}_{S_{N}})$
\STATE ETW $\leftarrow \max(f_{S_{N} \rightarrow S_{N+1}}^{-1}(\mathrm{ETW}),\mathrm{ETW}_{S_{N}})$
\IF{$\mathrm{ETW}_{S_N} > \mathrm{LTW}$}
\RETURN False
\ENDIF
\STATE \textbf{$\mathrm{BPS^{'}}$} $\leftarrow f_{S_{N} \rightarrow S_{N+1}}^{-1}( \mathrm{BPS^{'}})\cup \mathrm{BPS_{S_{N}\rightarrow S_{N+1}}}\cup \{\mathrm{ETW}_{S_{N}}  + \mathrm{W}_{S_{N}}, \mathrm{LTW}_{S_{N}} + \mathrm{W}_{S_{N}}\}$
\FOR{BP $\in$ \textbf{$\mathrm{BPS^{'}}$}}
\IF{$\mathrm{BP}\ \mathrm{not\ in}\ [\mathrm{ETW} + \mathrm{W}_{S_{N}}, \mathrm{LTW} + \mathrm{W}_{S_{N}}]$}
\STATE $\mathrm{BPS^{'}} \leftarrow \mathrm{BPS^{'}} \setminus{\mathrm{BP}}$
\ENDIF
\ENDFOR
\ENDFOR
\STATE \textbf{$\mathrm{BPS_{0}}$} $\leftarrow \mathrm{BPS^{'}}$
\STATE \textbf{forward propagation}
\STATE $\mathrm{min\_time\_cost} = +\mathrm{M}$
\FOR{$\mathrm{BP}_{0}$ $\in$ \textbf{$\mathrm{BPS_{0}}$}}
\FOR{$\mathrm{node}\ S_{N} \in (S_{1},S_{2}...,S_{v-1})$}
\STATE $\mathrm{BP}_{S_{N}} \leftarrow f_{S_{N-1} \rightarrow S_{N}} (\mathrm{BP}_{S_{N-1}})$
\ENDFOR
\IF{$\mathrm{BP}_{2n+1}-\mathrm{BP}_{0}<\mathrm{min\_cost}$}
\STATE $\mathrm{min\_time\_cost} \leftarrow \mathrm{BP}_{S_{N}}-\mathrm{BP}_{0}$
\STATE $\mathrm{departure\_time} \leftarrow \mathrm{BP}_{0}$
\ENDIF
\ENDFOR
\RETURN $\mathrm{departure\_time, min\_time\_cost}$
\end{algorithmic}
\end{algorithm}

\par First, note that for any arc \((i \rightarrow j)\), there exists a piece-wise linear mapping between the departure time at node \(i\) and the earliest ready time to departure at node \(j\). Any finite composition of piece-wise linear functions remains piece-wise linear. Therefore, when analyzing a single route $S = (S_{0},S_{1}...S_{v-1},S_{v})$, considering all these breakpoints is sufficient and necessary. In each iteration in backward propagation, we update the set of breakpoints for node $S_{N}$. It consists of three components: the time breakpoints caused by the time-dependent route $(S_{N} \rightarrow S_{N+1})$, the breakpoints inherited from the subsequent node $S_{N+1}$, and the earliest and latest time window of the current node plus the serving time $(\mathrm{ETW}_{S_{N}} + \mathrm{W}_{S_{N}}, \mathrm{LTW}_{S_{N}} + \mathrm{W}_{S_{N}})$. Then we eliminate all unfeasible and redundant breakpoints based on the updated time windows. After the backward propagation procedure, we get all feasible departure time breakpoints at depot $0$. Finally, we calculate the arrival time at depot $2n+1$ corresponding to each feasible departure time breakpoint. We can determine the optimal departure time by comparing the time differences corresponding to each breakpoint. Note that within a closed interval, the piece-wise linear function can only achieve the global optimum at the breakpoints, thus completing the evaluation procedure.

\subsection{Adaptive Large Neighborhood Search (ALNS)}\label{LNS_par}
\label{subsection:ALNS_details}
\par Based on the characteristics of the MTTD-MVRP, we have designed 5 removal operators and 4 insertion operators for the ALNS algorithm. In each iteration, dynamic perturbation strength determines the number of nodes to be removed and reinserted. Additionally, we employ adaptive roulette wheel selection to decide the order and extent of operators to apply. Each generated solution is then evaluated based on its cost and the number of new routes it contains to decide whether to retain it. The ALNS algorithm aims to perturb the incumbent solution significantly to avoid converging to the same local optimum solution after the local search (LS) procedure. Finally, we adaptively adjust the parameters influencing dynamic perturbation strength and the roulette wheel selection process based on the usage and performance of each operator.

\subsubsection{Removal operators}
\label{Removal operators}
\begin{itemize}
\item Distance-based Removal (DisR): The DisR operator defines the distance margin of a node \( i \) as \( Dis(S) - Dis(S^{-i}) \), where \( S \) represents the original solution, and \( S^{-i} \) represents the new solution obtained by removing node \( i \) from solution \( S \). The function \( Dis \) returns the total distance of the specific route. The DisR operator aims to remove the node with the maximum distance margin, effectively eliminating certain edges in the route to create a more compact and spatially efficient solution.

\item Duration-based Removal (DurR): The DurR operator functions similarly to the DisR operator but focuses on duration rather than distance. It defines the duration margin of a node \( i \) as \( Dur(S) - Dur(S^{-i}) \), where \( Dur \) returns the total duration of the route. The DurR operator removes the node with the largest duration margin, thereby eliminating nodes that cause conflicts with time windows. The removed node is then reinserted in a later phase to help minimize the total duration of the route.

\item Random Removal (RR): The RR operator randomly removes a node, introducing randomness to the LNS procedure. This randomness ensures the exploration of diverse solutions.

\item Shaw Removal (ShaR): The Shaw Removal operator is designed to remove nodes based on their compatibility with each other. It further removes adjacent nodes to create more space for reinsertion. We apply the closeness measure defined as \( CL(i,j) = \frac{dis(i,j)}{\bar{v}_{i,j}} + \alpha_{w} \times \mathrm{max}\{0,ETW_{j} - f_{i \rightarrow j}(LTW_{i} + W_{i})\} + \alpha_{v} \times \mathrm{max}\{0, f_{i \rightarrow j}(ETW_{i} + W_{i}) - LTW_{j}\} \) , where \( \bar{v}_{i,j} \) denotes the average speed traveling from \( i \) to \( j \), corresponding to the specific speed profile of arc \( (i \rightarrow j) \). Here, \( \alpha_{w} \) represents the waiting penalty parameter, scaled by the minimum possible waiting time duration for the vehicle, while \( \alpha_{v} \) represents the violation penalty parameter, scaled by the minimum possible wrap time for the vehicle \citep{pan2021hybrid}.

\item Segment Removal (SegR): The SegR operator removes a segment of consecutive nodes from the route. It can effectively break existing patterns and generate spatial and temporal gaps for later reinsertion.

\end{itemize}

\subsubsection{Insertion operators}
\begin{itemize}
\label{greedy reinsertion}
\item Greedy Insertion (GI): This operator selects a node from the pool of waiting nodes and inserts it into the route using a greedy algorithm, always aiming to minimize the cost in the new solution $S^{+i}$.

\item Regret-2 Insertion (R2I): The R2I insertion is an improved version of the greedy method with a forward-looking scheme. Firstly, defining $\Delta C_{i,r}$ to be the difference in time cost caused by inserting node i into the position r of the current solution. Then these values are sorted in ascending order of $\Delta C_{i,r}$ for each node i. Following that, we calculate the regret-2 value of node i as $R2V_{i} = \Delta C_{i,r_{(1)}} - \Delta C_{i,r_{(2)}}$, where $r_{(1)}$ and $r_{(2)}$ denote the inserting positions of node i with the lowest and second-lowest cost, respectively. Finally, we choose one node i with the largest $R2V_{i}$ and insert it into the best position $r_{1}$.

\item Regret-k Insertion (RkI): RkI works in the same way as R2I but with a generalized regret value. Here we give the formulation for general regret-k value: $RkV_{i} = \sum^{k}_{j=2}(\Delta W_{i,j} - \Delta W_{i,1})$. A greater k value setting makes the operator more forward-looking. However, considering more factors leads to a distorted determination. Typically, regret-2, regret-3, and regret-4 insertion operations are used in the ALNS. Note the RkI degenerates to the GI when k = 1. RkI has been shown able to produce high-quality solutions for complicated problems \citep{zhang2015memetic, lim2016HK_pdp}.

\item Random Insertion (RI): This operator picks a waiting node randomly and inserts it into the best position of the original route. The RI operator introduces randomness to the insertion procedure.

\item Segment Insertion (SI): This operator uses the regret-2 method to create a segment from the pool of waiting nodes. It then identifies the optimal position in the current solution to insert this segment, aiming to minimize the overall cost.

\end{itemize}

\subsubsection{Dynamic perturbation strength} \label{Dynamic perturbation strength}
The dynamic perturbation strength in ALNS determines the number of nodes removed from the current solution during each iteration. This perturbation strength is adjusted dynamically based on the difficulty of finding a new solution in previous iterations. To implement this, we introduce a counter $N-I$, which tracks the number of consecutive iterations without improvement. Every time ALNS fails to find a better solution, the counter increases by 1. When a new global optimum is discovered, the counter resets to 0. The number of nodes to be removed in the current iteration is calculated using $\min\{r_{\text{max}}, r_{\text{min}} + \alpha \times \text{N-I}\} \times n$, where $n$ is the total number of nodes, and $r_{\text{max}}, r_{\text{min}}$ are the maximum and minimum ratios of nodes that can be removed, respectively, and the parameter $\alpha$ determines how quickly the ratio of nodes to be removed increases with the $N-I$ counter.

\subsubsection{Roulette wheel selection}
We aim to allocate the total number of nodes to be removed, denoted as $X$ to each Removal operator based on its corresponding weight $w_{i}$. The number of nodes to be removed by $i_{th}$ operator is a random variable following $X_i \sim \text{Exp}\left(\lambda_i\right) \cdot \left(X - \sum_{j=1}^{i-1} X_j\right)$ where $\lambda_i = \frac{w_i}{\sum_{j=1}^{i-1} w_j}$. 
To ensure the order operators are applied is taken into account independently. We define a 2D list \( E_{(pos, i)} \) to represent the effectiveness of each operator at a certain position in the randomized sequence of LS operators. The LS operators are selected randomly from the remaining LS operator pool until it is empty. Let \( LS \) denote the set of all LS operators, and \( LS_1, LS_2, \ldots, LS_{z-1} \) be the first, second, ..., \( z \)-th LS operators selected. The remaining LS operators are \( LS \setminus \{LS_1, LS_2, \ldots, LS_{z-1}\} \). The \( z \)-th operator is then selected based on the probability \( P_{i} = \frac{E_{(z, i)}}{\sum_{j \in LS \setminus \{LS_1, LS_2, \ldots, LS_{i-1}\}} E_{(z, j)}} \).
During the initializing phase, we assign equal weights and effectiveness to $w_{i} and E_{(pos, i)}$. Once a new global optimum is discovered, the weight of operator $w_{i}$ increased by $\alpha_{1} \cdot R_{i}$, where $R_{i}$ represents the actual proportion of nodes removed by i operator. Additionally, the effectiveness $E_{(pos_{i}, i)}$ increased by $\beta_{1} \cdot R_{i}$, $pos_{i}$ indicate the position of i operator in the randomized sequence. If a new incumbent solution is found, another set of parameters \( (\alpha_2, \beta_2) \) is used to multiply \( R_i \), resulting in adjustments to \( w_i \) and \( E_{(pos_i, i)} \) respectively. We apply a similar mechanism to the repair process in ALNS as well.

\subsection{Local search}\label{LS_par}
\par Different from ALNS, the local search (LS) aims to find a local optimum in a small but promising area around a current existing solution. For the MTTD-MVRP, we customize several problem-specific LS operators.

\begin{itemize}
    \item \textbf{Swap}: The swap operator selects two nodes and swaps their positions in the original solution.
    \item \textbf{Relocate}: One node is selected and relocated into a new position in the original solution.
     \item \textbf{2-opt}: The 2-opt operator removes two edges in a single route and reconnects after reversing the order of nodes between them. This operator is aimed to eliminate crossing paths. 
     \item \textbf{2-opt*}: The 2-opt* is an extension of the 2-opt technique useful in multi-route problems, which removes one edge from each of two distinct routes and then reconnects the edges in an interlaced manner to create two new routes.
    \item \textbf{Transform}: This operator is uniquely designed for the MTTD-MVRP. It selects a node from the current solution. If the selected node is a delivery node \(v \in V_{d}\), it transforms the node into the corresponding SDL node \(v+n \in V_{s}\) without changing its position in the route. Conversely, if the selected node is an SDL node \(v+n \in V_{s}\), it transforms the node back into the corresponding delivery node \(v \in V_{d}\). 
    \item \textbf{Ant}: This operator draws inspiration from ACO. A pheromone matrix is constructed, and whenever other operators successfully modify the solution, the broken arcs and newly constructed arcs are recorded, with their corresponding pheromone levels updated. When the Ant operator is invoked, it exploits this pheromone information, prioritizing arcs with higher pheromone levels to modify the route by breaking and reconstructing arcs.
\end{itemize}

\subsubsection{Markov selection}\label{Markov selection}
\par This mechanism improves roulette wheel selection and traditional iterations used in tabu search. While roulette wheel selection considers the weight of each operator, it ignores the enhancing or diminishing effects between operators—some work better when preceded by specific others, while some lose effectiveness if executed consecutively. To address this, we maintain a transition probability matrix \( P_{x,y} \), which determines the probability of selecting the \( y^{th} \) operator after the \( x^{th} \). Given the current operator \( x \), we select the \( y^{th} \) operator with probability \( \frac{P_{x,y}}{\sum_{z \notin I}P_{x,z}} \), where \( I \) includes the ineffective operators. For instance, if the \( x^{th} \) operator fails to improve the solution, it is added to \( I \). Once an effective operator is found, the set \( I \) is reset to empty.

\par Each time a better solution is found using the \( y^{th} \) operator following the \( x^{th} \) operator, we reward \( P_{x,y} \) by adding \( \varepsilon_{+} \cdot \frac{\Delta_{obj}}{\lambda_{y}} \), where \( \varepsilon_{+} \) is a preset scale factor, \( \Delta_{obj} \) represents the change in the objective function, and \( \lambda_{y} \) is the real-time cost of the operator. If no improvement is found, we penalize it by subtracting \( \varepsilon_{-} \cdot \lambda_{y} \).

\subsubsection{Local search procedure}
\par We present the entire LS procedure in Algorithm \ref{LS}. The process begins by selecting an operator using the Markov selection process based on the previous effective operator \(e\), the set of ineffective operators \(I\), and the transition probability matrix \(P\)(Line \ref{LS_5}). The selected operator is applied to the current solution \(S'\), resulting in a temporary solution \(T\) (Line \ref{LS_1}). The best-feasible principle is employed to explore local optima. If the temporary solution \(T\) improves the cost function, the current solution \(S'\) is updated to \(T\), and the previous effective operator \(e\) is set to the current operator \(c\). Additionally, the set \(I\) is reset (Line \ref{LS_2}). If no improvement is found, the current operator \(c\) is added to the set \(I\). The adaptation process (Lines \ref{ad1} and \ref{ad2}) on \(P\) is described in the section \ref{Markov selection}. The LS loop terminates when all LS operators have failed to improve the solution (Line \ref{LS_4}).

\begin{algorithm}[htp]
\caption{Local Search Procedure}
\label{LS}
\begin{algorithmic}[1]
\REQUIRE $S'$
\STATE $N_{\mathrm{ls}}$ = 5, the total number of LS operators used
\STATE $e \leftarrow 0$, $I \leftarrow \emptyset$
\WHILE{$|I| \neq \mathrm{ls}$}
\label{LS_4}
\STATE $c \leftarrow \mathrm{Markov \ selection}(e,I,P)$
\label{LS_5}
\STATE $T\leftarrow  $ apply $c$-th operator to $S^{'}$ \label{LS_1}
\IF {$W(T) < W(S')$} 
\STATE $S' \leftarrow T$, $e \leftarrow c$, $I \leftarrow \emptyset$
\label{LS_2}
\STATE $\mathrm{Reward\ transition\ probability} P_{e,c}$
\label{ad1}
\ELSE
\STATE $I \leftarrow I \cup \{c\}$
\STATE $\mathrm{Penalize\ transition\ probability} P_{e,c}$
\label{ad2}
\ENDIF
\ENDWHILE
\RETURN $S'$
\end{algorithmic}
\end{algorithm}

\section{Computational experiments} 

\par In this section,  we conduct numerical experiments to evaluate the performance of our proposed meta-heuristic algorithm based on a set of newly designed test instances that were adapted from the test instances proposed by \cite{Dabia2013bp_tdvrptw} tailored for the MTTD-MVRP. We also provide extensive computational experiments to evaluate the performance of our meta-heuristic algorithm. 
Our algorithm is implemented in Python 3.9 and executed on a server equipped with an Intel(R) Xeon(R) Silver 4214R CPU at 2.40 GHz. For comparison, the theoretical solutions to the MILP model are computed using the Gurobi Optimizer (version 10.0), as outlined in Section \ref{MILP}.

Our experiments are organized into several key components. First, we adapt our algorithm to address DM-TDVRPTW instances by degenerating our problem setting to compare benchmark results from \cite{pan2021hybrid}. This comparison, along with a detailed discussion, is presented in Section \ref{Comparison on DM-TDVRPTW benchmark instances}. Second, we test our algorithm on small instances, comparing its performance against exact solutions provided by a MIP solver, and demonstrate its capability on larger instances for the MTTD-MVRP (Section \ref{Large-Scale MTTD-MVRP Testing}). Third, we examine the effectiveness of the Markov selection mechanism by selectively enabling or disabling it to assess its contribution to algorithm performance. Finally, we explore the impact of varying SDL service parameters to get some management insights.

\subsection{Comparison on DM-TDVRPTW benchmark instances} \label{Comparison on DM-TDVRPTW benchmark instances}
\par To evaluate the effectiveness of our algorithm, we perform a comparative analysis against the leading ALNS-TS algorithm introduced by \cite{pan2021hybrid}, which was generated by solving the DM-TDVRPTW instances proposed by \cite{Dabia2013bp_tdvrptw}. which are derived from the well-known Solomon dataset. These instances include customer sets of 25, 50, and 100 (labeled as T-25, T-50, and T-100, respectively) and are categorized into three groups based on the geographic distribution of customers: random (R), clustered (C), and a mixed distribution (RC). Furthermore, two types of time windows are considered: Type 1, tight time windows, and Type 2, wider time windows.

To evaluate the effectiveness of our algorithm, we conduct a comparative analysis against the advanced ALNS-TS algorithm proposed by \cite{pan2021hybrid}. The benchmark was constructed using the DM-TDVRPTW instances introduced by \cite{Dabia2013bp_tdvrptw}, derived from the widely recognized Solomon dataset. These instances consist of customer sets of 25, 50, and 100, denoted as T-25, T-50, and T-100, respectively. They are further categorized into three groups based on the geographic distribution of customers: random (R), clustered (C), and mixed distribution (RC). Additionally, two types of time windows are considered: Type 1 with tighter constraints and Type 2 with broader flexibility.

Tables \ref{T-25}, \ref{T-50}, and \ref{T-100} summarize the comparison results, reporting the average best cost presented in \cite{pan2021hybrid} ($Cost_{A}$), the average best cost achieved by our ALNS-LS algorithm ($Cost_{LS}$), and the percentage gap calculated as \((Cost_{LS} - Cost_{A}) / Cost_{A} \times 100\). 

\textbf{Conclusion:} Our algorithm establishes 7 new best-known solutions out of 12 total instances in the R1 100-customer group and 4 new best-known solutions out of 8 total instances in the RC1 100-customer group. ALNS-LS demonstrates competitive performance across Type 1 instances and successfully identifies several new best-known solutions.

\begin{table}[h!]
\centering
\caption{T-25}
\begin{tabular}{lcccccc}
\hline
\label{T-25}
Grp &  & ALNS-TS $Cost_{A}$ &  & ALNS-LS $Cost_{LS}$ &  & Gap (\%)  \\ 
\hline
C1  &  & 24507.86           &  & 24507.86            &  & -       \\
R1  &  & 6519.09            &  & 6519.09             &  & -       \\
RC1 &  & 5784.09            &  & 5784.09             &  & -       \\
C2  &  & 24840.22           &  & 24840.22            &  & -       \\
R2  &  & 6191.81            &  & 6191.81             &  & -       \\
RC2 &  & 6455.35            &  & 6455.35             &  & -       \\
\hline
\end{tabular}
\end{table}

\begin{table}[h!]
\centering
\caption{T-50}
\begin{tabular}{lcccccc} 
\hline
\label{T-50}
Grp &  & ALNS-TS $Cost_{A}$ &  & ALNS-LS $Cost_{LS}$ &  & Gap (\%)  \\ 
\hline
C1  &  & 48487.36           &  & 48487.36            &  & -       \\
R1  &  & 11681.05           &  & 11681.05            &  & -       \\
RC1 &  & 11662.50           &  & 11659.69            &  & -0.024  \\
C2  &  & 48852.11           &  & 48852.11            &  & -       \\
R2  &  & 10873.28           &  & 10873.28            &  & -       \\
RC2 &  & 11619.13           &  & 11620.91            &  & 0.015   \\
\hline
\end{tabular}
\end{table}

\begin{table}[h!]
\centering
\caption{T-100}
\begin{tabular}{lcccccc} 
\hline
\label{T-100}
Grp &  & ALNS-TS $Cost_{A}$ &  & ALNS-LS $Cost_{LS}$ &  & Gap (\%)  \\ 
\hline
C1  &  & 97473.19           &  & 97485.60            &  & 0.013    \\
R1  &  & 20187.11           &  & 20175.47            &  & -0.058   \\
RC1 &  & 21434.01           &  & 21432.58            &  & -0.007   \\
C2  &  & 95764.18           &  & 95776.62            &  & 0.013    \\
R2  &  & 18564.13           &  & 18568.21            &  & 0.022    \\
RC2 &  & 22789.01           &  & 22795.59            &  & 0.029    \\
\hline
\end{tabular}
\end{table}

\subsection{Evaluation on MTTD-MVRP instances}
\par To evaluate the proposed algorithm, we generate new problem-specific MTTD-MVRP instances by adapting the DM-TDVRPTW dataset. These instances are labeled as 10-MT, 25-MT, 50-MT, and 100-MT, representing instances with 10, 25, 50, and 100 customers, respectively. Each customer is associated with a Self-Delivery Locker (SDL) near their residence, mostly offering a more flexible time window. we also incorporate an alternative vehicle type, Autonomous Electric Vehicles (AEVs), which are constrained by a maximum driving distance and feature a comparatively lower fixed cost than traditional vehicles. Additionally, customers are assigned predefined probabilities of accepting SDL or AEV services. All MTTD-MVRP instances are publicly available at \url{https://github.com/vega-arclight/MTTD-MVRP-Instances.git.}

\subsubsection{Comparison with exact solutions on small instances}
\par To validate the accuracy and efficiency of the proposed meta-heuristic algorithm, we utilize Gurobi’s MILP solver to obtain exact solutions for small-scale MTTD-MVRP instances. By comparing the solutions provided by the meta-heuristic to the exact results, we evaluate its ability to approximate optimal solutions within a reasonable computational time. The results of this comparison are detailed in Table \ref{comparison on small-scale MTTD-MVRP instances}.

\begin{table}[h!]
\centering
\caption{comparison on small-scale MTTD-MVRP instances}
\label{comparison on small-scale MTTD-MVRP instances}
\begin{tabular}{m{0.15\linewidth} c m{0.25\linewidth} c c m{0.25\linewidth} c}
\hline
\multirow{2}{*}{Grp, 10\_MT} &  & \multicolumn{2}{c}{\textbf{Gurobi}} & & \multicolumn{2}{c}{\textbf{ALNS-LS}} \\ 
\cline{3-4} \cline{6-7}
                           &  & $\mathrm{Avg\ C_{best}}$ & $\mathrm{Avg\ T}$ & & $\mathrm{Avg\ C_{best}}$ & $\mathrm{Avg\ T}$ \\ 
\hline
R                          &  & 9985.83 & 61.04 & & 9985.83 & 2.52 \\
C                          &  & 3349.27 & 48.79 & & 3349.27 & 6.64 \\
RC                         &  & 3446.91 & 76.61 & & 3446.91 & 8.92 \\ 
\hline
\end{tabular}
\end{table}

\par We provide the average best costs obtained from Gurobi and our heuristic algorithm. For relatively small datasets, the heuristic algorithm results match exactly with the MILP solver's results. Additionally, the ALNS-LS algorithm demonstrates superior computational efficiency. Furthermore, even for the 25-MT instance, Gurobi fails to produce a satisfactory solution within a practical time limit.

\subsubsection{Large-Scale MTTD-MVRP Testing} \label{Large-Scale MTTD-MVRP Testing}
\par To demonstrate the scalability and effectiveness of the proposed algorithm, we test it on the following large-scale MTTD-MVRP instances, including the 25-MT, 50-MT, and 100-MT datasets. The MTTD-MVRP instances are more complex than standard VRP variants, with double the nodes due to SDLs, dual time windows per customer, and varied customer preferences for AEVs and SDL services, all adding layers of difficulty to the routing optimization. The experiments aim to assess the algorithm's ability to handle large datasets while maintaining solution quality efficiently. The ALNS-LS is executed 5 runs for each instance, with a maximum runtime of 20 seconds for 25-MT, 300 for 50-MT, and 5500 seconds for 100-MT, and the results are summarized in Table \ref{Large scale MT instance}. 

\begin{table}[h!]
\centering
\caption{Large scale MT instance}
\renewcommand{\arraystretch}{1.2} 
\label{Large scale MT instance}
\begin{tabular}{llcccc} 
\toprule
Size & Grp & $C_{avg}$ & $C_{best}$ & Gap (\%) & $T_{best}$ \\ 
\midrule
\multirow{3}{*}{25\_MT} 
    & C   & 25,611.66  & 25,611.66  & -      & 3.31       \\
    & R   & 7,494.12   & 7,494.12   & -      & 4.46       \\
    & RC  & 7,405.37   & 7,405.37   & -      & 3.76       \\ 
\midrule
\multirow{3}{*}{50\_MT} 
    & C   & 50,546.63  & 50,546.63  & -      & 60.03      \\
    & R   & 14,166.16  & 14,163.39  & 0.0195   & 259.59     \\
    & RC  & 13,765.86  & 13,757.57  & 0.0602   & 281.82     \\ 
\midrule
\multirow{3}{*}{100\_MT} 
    & C   & 101,579.02 & 101,579.02 & -      & 591.06     \\
    & R   & 25,620.86  & 25,585.28  & 0.1389   & 5,177.69   \\
    & RC  & 26,370.49  & 26,359.56  & 0.0414   & 4,344.16   \\ 
\bottomrule
\end{tabular}
\end{table}

\subsection{Evaluation of algorithmic components}
\par In this section, we evaluate the effectiveness of each key component of our hybrid meta-heuristic ALNS-LS algorithm. The algorithm consists of three main components: large-scale exploration and local optima escape using the ALNS procedure, intensive local refinement through the LS procedure, and the dynamic selection and connection of operators enabled by the Markov selection mechanism.
To assess the performance of these components, we conduct experiments using the T-100-MT and T-100 instances. Each test is run ten times per instance to ensure statistical reliability. We focus on quantifying each component's contribution to the overall solution quality and computational efficiency. Detailed results are presented in the following.

\subsubsection{Impact of ALNS and LS Procedures}
To examine the individual contributions of the ALNS and LS procedures, we conduct experiments by disabling each component separately and evaluating their effects on the final results. For comparison, three algorithmic variants are considered:

\begin{itemize} 
\item \textbf{ALNS-LS}: The complete original algorithm, incorporating both ALNS and LS procedures. 
\item \textbf{LS}: A variant where the ALNS procedure is disabled. \item \textbf{ALNS}: A variant where the LS procedure is disabled. \end{itemize}

\par We evaluated the performance of three algorithmic variants ALNS-LS, ALNS, and LS on the MT-100 instances, running each variant five times per instance. The best cost and the corresponding runtime were recorded, and the average resulting cost (\( C_{avg} \)) was computed. The termination condition was set to 200 seconds of no further improvements in the solution. Table \ref{Comparison of ALNS-LS, ALNS, and LS} summarizes the results, with the Gap (\%) calculated based on the difference between \( C_{best} \) and the larger value among the results.

\par we test these variants on MT-100, running them 5 times on each instance, recording the best cost and corresponding time taken, and calculating the average resulting cost. We set the termination condition as once the algorithm cannot find a better solution in 200 seconds. Table \ref{Comparison of ALNS-LS, ALNS, and LS} presents the detailed statistics, and the Gap is calculated based on the difference between $C_{best}$ over the larger value. This result clearly demonstrates the pros and cons of ALNS and LS. ALNS performed a table on each group, their $C_{avg}$ does not deviate from $C_{best}$ too much, but it was hard to refine the final solution into a better one, resulting in a neglectable gap. LS aims to refine the final solution, leaking the ability to escape from the local optima far away, so it takes much less time to terminate and the result is sensitive based on the initial solution and some randomness resulting in an unpredictable $C_{avg}$ and sometimes it may approach a satisfactory solution. 



\begin{table}[htbp]
\centering
\caption{Comparison of ALNS-LS, ALNS, and LS}
\label{Comparison of ALNS-LS, ALNS, and LS}
\label{tab:comparison-ALNS-LS-ALNS-LS}

\begin{tabular}{lll S[table-format=6.2] S[table-format=6.2] S[table-format=6.2] c}
\toprule
  Var & & Grp & {$C_{\text{avg}}$} & {$C_{\text{best}}$} & {$T_{\text{best}}$} & {Gap (\%)} \\
\midrule
\multirow{3}{*}{ALNS-LS} & & C  & 101582.74 & 101579.02 &  591.06 & -    \\
                         & & R  &  25603.51 &  25585.28 & 5177.69 & -    \\
                         & & RC &  26367.84 &  26359.56 & 4344.16 & -    \\
\midrule
\multirow{3}{*}{ALNS}    & & C  & 107128.94 & 106942.28 &  348.84 & 5.02 \\
                         & & R  &  27538.17 &  27382.52 & 3989.71 & 6.56 \\
                         & & RC &  27395.13 &  27271.63 & 3718.89 & 3.35 \\
\midrule
\multirow{3}{*}{LS}      & & C  & 105319.76 & 102895.81 &  253.92 & 1.28 \\
                         & & R  &  28159.01 &  26818.97 &  393.65 & 0.87 \\
                         & & RC &  27291.47 &  26914.57 &  211.78 & 2.00 \\
\bottomrule
\end{tabular}
\end{table}

The statistics illustrate the performance of ALNS and LS. ALNS exhibits consistent performance across all groups, with a negligible difference between \( C_{avg} \) and \( C_{best} \). However, its inability to refine solutions further results in a larger overall gap when compared to ALNS-LS. LS, on the other hand, is good at fine-tuning solutions. However, it struggles to escape suboptimal local minima, making its performance highly dependent on the initial solution and inherent randomness and introducing variability in \( C_{avg} \). In some instances, LS approaches satisfactory solutions but lacks stability.

\subsubsection{Impact of the Markov Selection Mechanism}

\par As noted in \cite{soylu2015general}, the order in which operators are called significantly influences the final results. Therefore, it is valuable to evaluate how the Markov selection mechanism self-adapts the operator sequence and enhances the overall algorithmic performance. To assess this impact, we conduct experiments by applying the LS procedures to a single instance, running them 50 times with and without the Markov selection mechanism. During these tests, we track the average number of invocations for each LS operator and calculate their success ratio, defined as the proportion of invocations that lead to an improvement over the incumbent solution.

\par
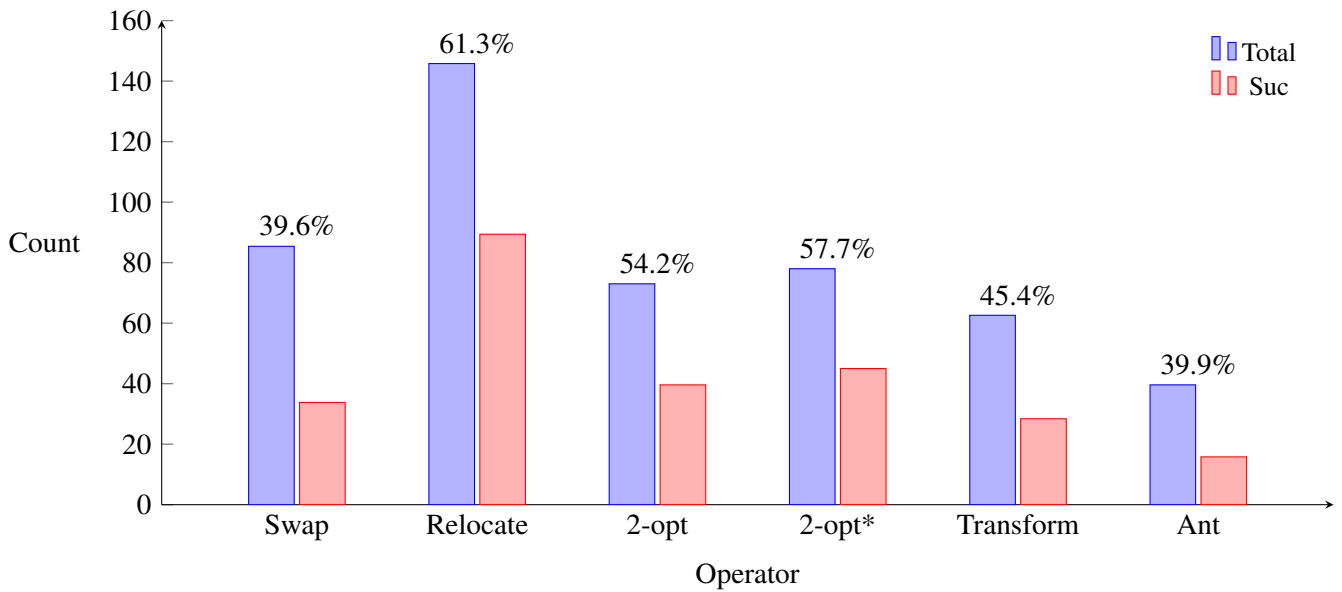
\begin{figure}[h!]
\centering
\label{LS operators performance with Markov Selection}
\begin{tikzpicture}
    \begin{axis}[
        ybar,
        bar width=0.6cm, 
        width=\textwidth,
        height=8cm,
        ymin=0,
        ymax=160,
        ylabel={Count},
        xlabel={Operator},
        symbolic x coords={Swap, Relocate, 2-opt, 2-opt*, Transform, Ant},
        xtick=data,
        legend style={at={(0.98,0.98)}, anchor=north east, draw=none, fill=none, font=\small},
        axis lines=left, 
        enlarge x limits=0.15, 
        enlarge y limits=false, 
        tick style={line width=0.5pt}, 
        every axis x label/.style={at={(0.5,-0.1)}, anchor=north}, 
        every axis y label/.style={at={(-0.1,0.5)}, anchor=south}, 
        xtick align=inside, 
        ytick align=inside, 
        xtick style={draw=none}, 
        ytick style={line width=0.5pt} 
    ]
        \addplot coordinates {(Swap,85.4) (Relocate,145.8) (2-opt,73) (2-opt*,78) (Transform,62.6) (Ant,39.6)};
        \addplot coordinates {(Swap,33.8) (Relocate,89.4) (2-opt,39.6) (2-opt*,45) (Transform,28.4) (Ant,15.8)};
        \legend{Total, Suc}

        \node[anchor=south] at (axis cs:Swap,85.4) {39.6\%};
        \node[anchor=south] at (axis cs:Relocate,145.8) {61.3\%};
        \node[anchor=south] at (axis cs:2-opt,73) {54.2\%};
        \node[anchor=south] at (axis cs:2-opt*,78) {57.7\%};
        \node[anchor=south] at (axis cs:Transform,62.6) {45.4\%};
        \node[anchor=south] at (axis cs:Ant,39.6) {39.9\%};
    \end{axis}
\end{tikzpicture}
\caption{LS operators performance with Markov Selection}
\end{figure}

\begin{figure}[h!]
\centering
\label{LS operators performance without Markov Selection}
\begin{tikzpicture}
    \begin{axis}[
        ybar,
        bar width=0.6cm, 
        width=\textwidth,
        height=8cm,
        ymin=0,
        ymax=140,
        ylabel={Count},
        xlabel={Operator},
        symbolic x coords={Swap, Relocate, 2-opt, 2-opt*, Transform, Ant},
        xtick=data,
        legend style={at={(0.98,0.98)}, anchor=north east, draw=none, fill=none, font=\small},
        axis lines=left, 
        enlarge x limits=0.15, 
        tick style={line width=0.5pt}, 
        every axis x label/.style={at={(0.5,-0.1)}, anchor=north}, 
        every axis y label/.style={at={(-0.1,0.5)}, anchor=south} 
    ]
        \addplot coordinates {(Swap,78.6) (Relocate,84.2) (2-opt,81.2) (2-opt*,69.6) (Transform,94) (Ant,86.4)};
        \addplot coordinates {(Swap,24.4) (Relocate,43.8) (2-opt,38.6) (2-opt*,40) (Transform,34) (Ant,17.8)};
        \legend{Total, Suc}

        \node[anchor=south] at (axis cs:Swap,78.6) {31.04\%};
        \node[anchor=south] at (axis cs:Relocate,84.2) {52.02\%};
        \node[anchor=south] at (axis cs:2-opt,81.2) {47.54\%};
        \node[anchor=south] at (axis cs:2-opt*,69.6) {57.47\%};
        \node[anchor=south] at (axis cs:Transform,94) {36.17\%};
        \node[anchor=south] at (axis cs:Ant,86.4) {20.60\%};
    \end{axis}
\end{tikzpicture}
\caption{LS operators performance without Markov Selection}
\end{figure}
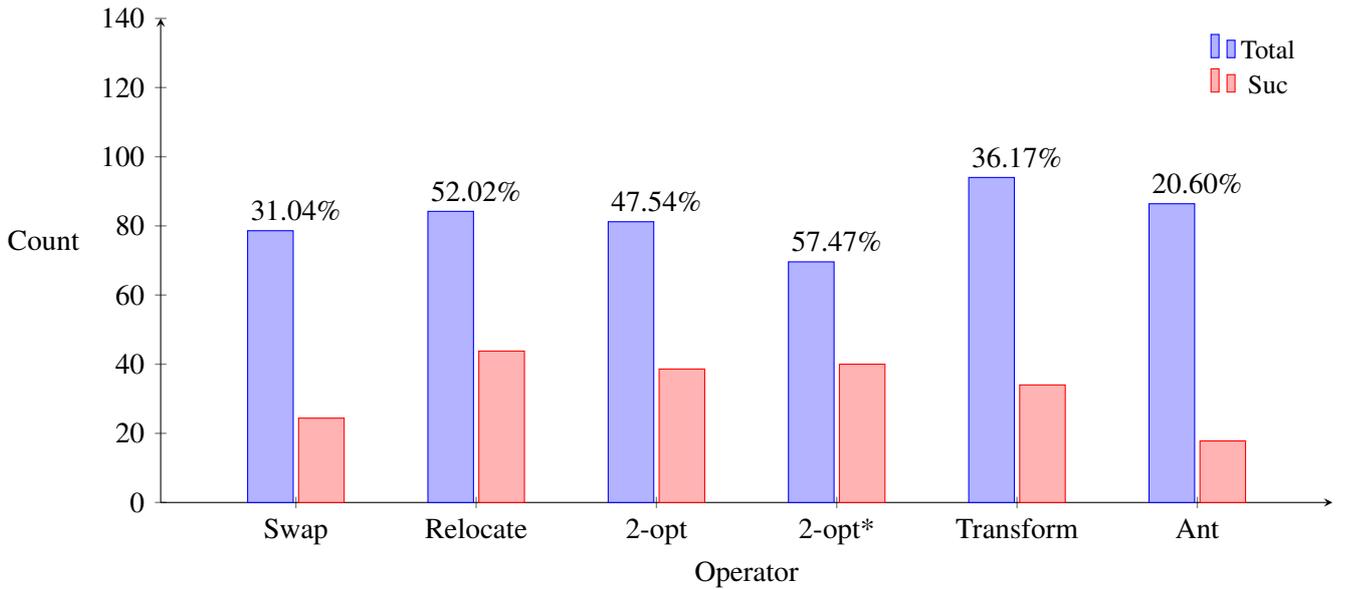

 The results of this analysis are illustrated in Figure \ref{LS operators performance with Markov Selection} and \ref{LS operators performance without Markov Selection}. It is obvious that Markov's Selection boosts the successful ratio of each operator, instead of using each operator averagely, Markov's Selection adapted to use more operators with better performance and boost the operators with fewer invocations.

\par The results are presented in Figures \ref{LS operators performance with Markov Selection} and \ref{LS operators performance without Markov Selection}. This demonstrates that the Markov selection mechanism improves the success ratio of each operator. Unlike a uniform invocation approach, the Markov selection mechanism dynamically adapts by prioritizing operators with better performance while also boosting the performance of less-invoked operators.

\subsection{Management Insights: Effects of SDL Spatial Distribution}
\par To derive actionable management insights, this subsection investigates the influence of the spatial distribution of SDLs on final costs and logistics performance. Specifically, we analyze how the density and proximity of SDLs impact delivery efficiency, the resultant cost, and the potential implications to the last-mile distances from SDLs to customer homes.
\subsubsection{Test Dataset Generation}
A new dataset was developed based on the RC 100-MT instances to assess the impact of SDL density on logistics performance. For each original instance,  all settings were kept unchanged except for the SDL coordinates, and seven variations were created labeled as 100-MT-1, 100-MT-2, ..., 100-MT-7, representing different levels of SDL density. For example, 100-MT-1 corresponds to the instance with the least compact SDL distribution, where SDLs are widely dispersed and shared usage is minimal. In contrast, 100-MT-7 represents the instance with the most compact SDL distribution, characterized by customers in close proximity frequently sharing SDLs, with some customers experiencing longer last-mile distances between their homes and the corresponding SDL.

\begin{figure}[htbp]
\centering
\includegraphics[width=1.05\textwidth]{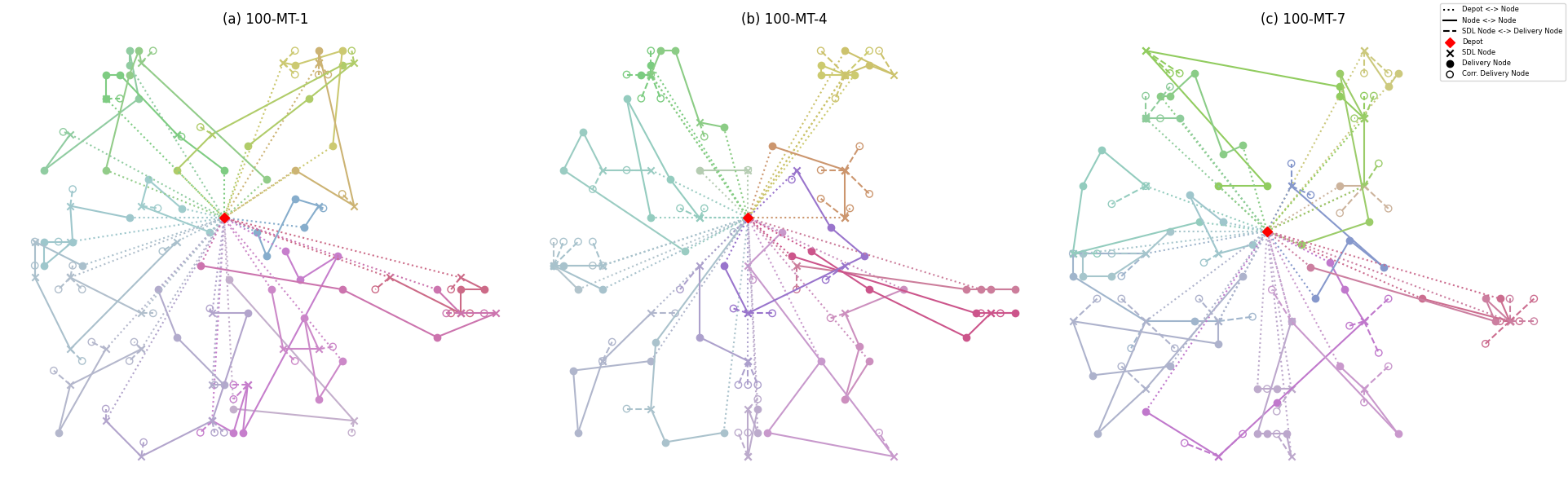}
\caption{SDL Density Variations Visualization}
\label{SDL Density Variations Visualization}
\centering
\end{figure}

Figure \ref{SDL Density Variations Visualization} illustrates three variations in SDL density. Instance (a) depicts the least dense distribution, characterized by widely dispersed SDLs with infrequent shared usage. Instance (b) represents a moderately dense distribution. Instance (c) illustrates the densest scenario, where SDLs are highly concentrated and frequent shared usage occurs among customers within neighborhoods. If an SDL is utilized for a customer, a dashed line is used to connect between the customer and the corresponding SDL. From left to right, there is a clear trend of increasing SDL compactness, with more customers sharing a common neighboring SDL. This shift reflects a potentially more efficient logistics system due to reduced total travel distances for vehicles. However, it also introduces the challenge of longer last-mile distances for individual customers in some cases.

\subsubsection{Numerical Experiment and Management Insights}
\begin{figure}[htbp]
\centering
\includegraphics[width=1.05\textwidth]{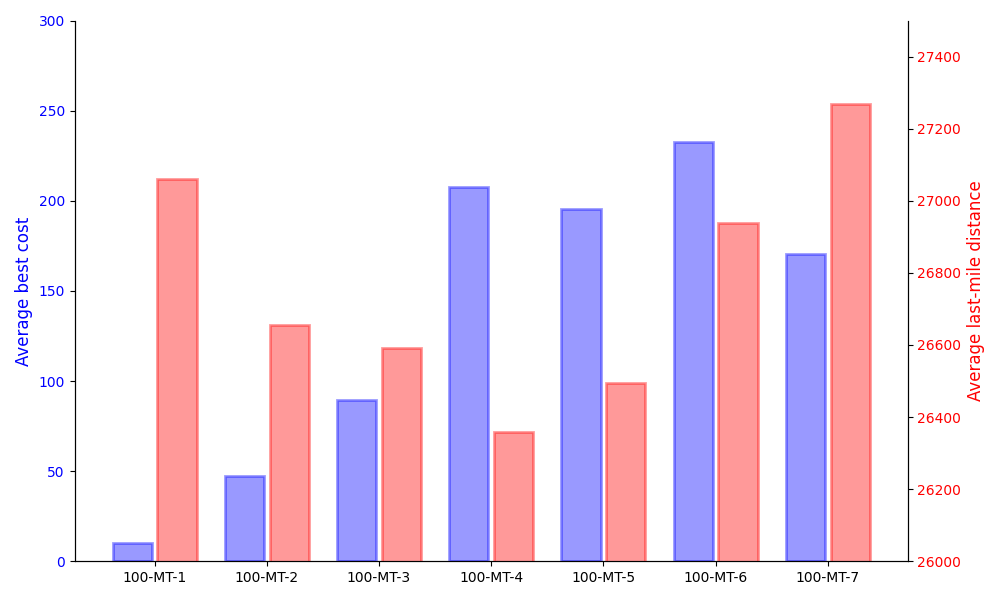}
\caption{Comparison of Average Best Cost and Last-Mile Distance Across Different SDL Distributions}
\label{SDL_exp.png}
\centering
\end{figure}

To evaluate the effects of SDL distribution on logistics performance, numerical experiments were conducted for the seven instances (100-MT-1 to 100-MT-7). Two key evaluation metrics were considered: the average best cost and the average last-mile distance. Each instance was executed 10 times, and for every run, the best cost and the corresponding last-mile distance were recorded. These values were then averaged across all instances with the same SDL density. Graph \ref{SDL_exp.png} presents the results, while detailed data are summarized in Table \ref{SDL_exp_table.png}, demonstrating how the average best cost and last-mile distance vary across different SDL distributions.

\begin{table}[h!]
\centering

\begin{tabular}{lccccccc} 
\hline
\label{SDL_Summary}
Metric                     & MT-100-1 & MT-100-2 & MT-100-3 & MT-100-4 & MT-100-5 & MT-100-6 & MT-100-7 \\ 
\hline
\textcolor{red}{Last-Mile} & 10.15    & 47.19    & 89.25    & 207.51   & 195.67   & 232.67   & 170.63   \\
\textcolor{blue}{Cost} & 27059.08 & 26654.93 & 26592.42 & 26359.56 & 26494.85 & 26937.71 & 27269.67 \\
\hline
\end{tabular}
\caption{Detailed Data Corresponding to Figure \ref{SDL_exp.png}}
\label{SDL_Summary}
\end{table}

\par As SDL density increases, vehicle routing efficiency improves due to reduced travel distances, resulting in lower average costs, particularly in moderately dense SDL distributions such as 100-MT-4 and 100-MT-5. However, excessively compact SDL distributions, as seen in 100-MT-6 and 100-MT-7, lead to diminishing returns, 
\section{Conclusion} 
\label{Section:Conclusions}
This study introduces the MTTD-MVRP, a novel problem that integrates AEVs and SDLs into last-mile logistics. By addressing time-dependent travel speeds, customer-specific preferences, and vehicle-specific constraints, we provide a robust and realistic model that aligns with modern urban logistics demands. 

We propose a hybrid ALNS-LS framework, equipped with tailored operators with a Markov selection mechanism, which effectively balances exploration and exploitation during the search process. Computational experiments validate the superior performance of our algorithm, achieving competitive results across large-scale MTTD-MVRP instances and improving upon benchmarks from Duration-Minimizing TDVRPTW problems. Notably, our method identifies new best-known solutions for several instances, demonstrating its ability to tackle both complexity and scale efficiently.

The findings suggest promising directions for integrating AEVs and SDLs into logistics, highlighting their potential to reduce costs and enhance delivery flexibility. Future research could extend this work by considering multi-vehicle type and multi-objective optimization for balancing cost, environmental impact, and customer satisfaction. Furthermore, incorporating stochastic elements such as demand uncertainty or dynamic time windows could enhance the realism and applicability of the proposed framework. By addressing these challenges, this study lays a foundation for innovative approaches to hybrid vehicle routing in urban logistics.


\section*{Acknowledgements}



\bibliographystyle{model5-names} 

\bibliography{tddarp}

\newpage
\section*{Appendix}
This is the appendix section.

\end{document}